# Towards Robotic Knee Arthroscopy: Multi-Scale Network for Tissue-Tool Segmentation

***Shahnewaz Ali, Prof. Ross Crawford, Dr. Frederic Maire, Assoc. Prof. Ajay K. Pandey***

School of Electrical Engineering and Robotics, Queensland University of Technology, Gardens Point, Brisbane, QLD 4001, AUSTRALIA.

## Abstract:

**Tissue awareness has a great demand to improve surgical accuracy in minimally invasive procedures. In arthroscopy, it is one of the challenging tasks due to surgical sites exhibit limited features and textures. Moreover, arthroscopic surgical video shows high intra-class variations. Arthroscopic videos are recorded with endoscope known as arthroscope which records tissue structures at proximity, therefore, frames contain minimal joint structure. As consequences, fully conventional network-based segmentation model suffers from long- and short- term dependency problems. In this study, we present a densely connected shape aware multi-scale segmentation model which captures multi-scale features and integrates shape features to achieve tissue-tool segmentations. The model has been evaluated with three distinct datasets. Moreover, with the publicly available polyp dataset our proposed model achieved 5.09 % accuracy improvement.**

# I. Introduction

Vision based navigation and localization are the fundamental problems of determining pose and location of the camera and subsequently drive robots towards the next direction. These are the crucial tasks for a vision based autonomous system. Several computer vision methods [1,2] are applied so far. Among others structure-based methods are gaining traction where semantic 3D maps represent the scene database, therefore, the camera pose is estimated through best matching of the query scene to the database. On the other hand, collision avoidance vision-based robot navigation systems rely on the semantic meaning of the scene to generate the next moving trajectory [3]. Hence, segmented maps are essential for robotics. In this study, surgical scene understanding (instance segmentation) is addressed to the context of robot assisted knee arthroscopy- the minimally invasive surgical (MIS) procedure to treat knee joint ailments. However, conventional knee arthroscopy also gets the identical benefits.
During the knee arthroscopy, a miniaturized camera known to as arthroscope and tools are introduced into the knee cavity through small incisions. It offers several benefits to patients, for instance, less surgical trauma, minimum tissue and blood loss, and quick recovery time. However, indirect vision of the surgical site, diminished hand-eye coordination, reduced field of view (FoV), lack of perception and haptic feedback, and limited access to the operating space are the major drawbacks [4], therefore, unintentional tissue damage is common to knee arthroscopy [5]. Additionally, lack of tissue awareness associated with the knee arthroscopy causes prolonged learning curves [4]. During the course of knee arthroscopy, surgeons often lose their confidence to recognize some tissue structures such as meniscus. As a matter of fact, they can only identify tissue-type femur with high confidence when only static video frames are provided. In order to increase their confidence, very often they used to recourse the arthroscopy from a known landmark structure. This additional navigation also increases the risk of unintentional tissue damage.

Apparently, an automatic tissue-tool segmentation model has a clear demand to mitigate these drawbacks. On the other side, segmented maps are essentials for robot-assisted knee arthroscopy - a means to localize and navigate camera-tools inside the cramped and confined knee cavity. However, it is a highly challenging task even for the skilled clinicians to segment tissue structure from static frames and often they use tracking information from video scenes. There are several factors associated with knee arthroscopy that makes the segmentation process extremely challenging. Visual representation exhibits highly texture-less structure and subsequently provides lack of discriminative image features. Moreover, close proximity between tissue structure and camera with narrow FoV make the segmentation process more complicated where in many frames only small fractions of joint structure are accessible with minimum contextual information. Furthermore, to create additional space for tools and to remove unwanted presence of blood, the saline water irrigation system often used in arthroscopy. This and other several environmental factors related to the underwater robotics such as motion, attenuation, and illumination conditions are the primary sources of poor imaging condition which limit the quality of video frames. Most standard factors are noise, debris, blur, saturation, and obscure frames. Besides that, it is hard to generalize the representation of each tissue type due to their significant structural irregularity and appearances caused by aging, degeneration, injury etc.

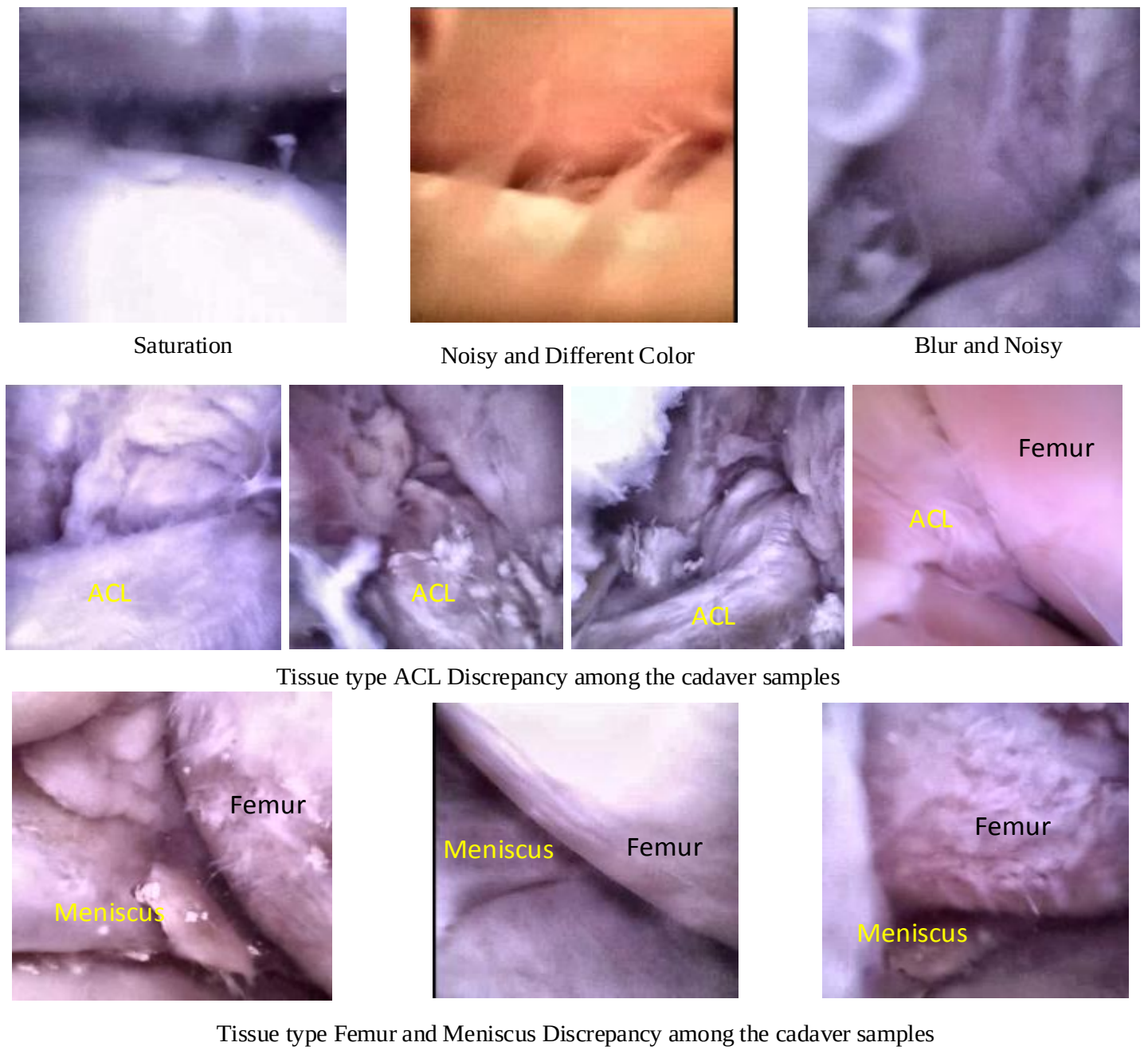

**Fig.1.** Representation of arthroscopic frames. The first row represents frames which are compromised by several attenuation factors such as illumination, saturation, noise and blur. The second row represents inconsistency of tissue type ACL in their appearance, structure and shapes. The third row represents inconsistency of tissue type femur (cartilage) and meniscus due to tissue degeneration.

In this study, we have proposed a deep learning architecture to achieve automatic tissue-tool segmentation from arthroscopic video frames. The proposed model is built on top of UNet, a deep learning architecture for biomedical image segmentation [6]. Our proposed network segments five key tissue-object structures: The femur, tibia, Anterior Cruciate Ligament (ACL), meniscus, and patella. It also segments any surgical-tools present in the surgical scene. The summary of our contributions in this study are as follows;

1. We proposed a deep learning architecture to achieve automatic tissue-tool segmented maps of arthroscopic video frames. To the best of our knowledge, this is the first time we are proposing a model for arthroscopy to achieve a complete scene-segmented map for key tissue structures and tools.
2. Our architecture addresses long- and short-range dependency problems of fully convolutional (FCN) based networks, explicitly applicable for UNet-like architecture. Moreover, the proposed network architecture minimizes information loss thus enabling strengthened information propagation at multi-scale. We also used spatial and channel-wise attention mechanisms, as well as global context block to embedded context aware learning. Additionally, the architectural design preserves salient features during the multi-scale feature extraction and dimensional alteration phases.
3. We integrated a secondary network branch as a shape feature extractor, which enables us to extract shape information of key tissue structures. Our proposed two-branch, multi-scale, densely connected, shape and context-aware network for segmentation higher accuracy the arthroscopic dataset. Moreover, we also received 5.09% accuracy gain when tested on the publicly available dataset for polyp segmentation.
4. We explicitly address data imbalance problems in the context of knee arthroscopy.

# II. Related work

With the success of deep learning techniques in computer vision, recently it has been successfully applied to segment various modality of biomedical images. More specifically, UNet architecture and its variants received groundbreaking success in laparoscopic intraoperative imaging techniques [6-7]. However, regarding arthroscopy a very scarce literature exists and most of the literature address different modality of images such as MRI [8-9], which are different from arthroscopic frames. In progress of robot assisted knee arthroscopy and to full-fill the demand of an extended intra-operative knee joint perception, Y. J. [4] applied UNet and UNet++ architecture to segment mainly four key tissue structures, namely Femur, Tibia, ACL and Meniscus. He received the highest average dice score of 0.79, 0.50, 0.51, 0.48 for Femur, Tibia, ACL, and Meniscus using the UNet++. The major limitations are originated from the dataset while the improved skip connections of UNet++ slightly improved the segmentation accuracy. Both networks suffered from lack of visual cues such as texture and feature. S.A et. al. [10] used classic UNet architecture and multispectral images reconstructed from RGB video frames to segment three tissue types. His approach slightly improves the segmentation accuracy, but the approach is strongly limited to the imaging conditions. M.S et.al. [11] segments arthroscopic frames to identify possible instrument gaps to autonomously manipulate legs and navigate instruments inside the cramped cavity. His work is restricted to binary segmentation intended to find joint gaps. To navigate instruments safely and to avoid unintended tissue damage from visual data, it is essential to recognize tools and the appearing tissue structures. An accurate tissue-tool segmentation procedure for arthroscopic video frames remains an open challenge.

# III. Methodology

## A. Dataset

This work is evaluated against two distinct knee arthroscopy data sets. A new stereo arthroscope is developed in our laboratory at Queensland University of Technology which supports relatively wider FoV with respect to standard arthroscopes. Moreover, standard arthroscope is a metallic tool having sharp edges. It supports 30 or 70-degree FoV and spot surgical sites from an angle. It diminishes the hand-eye coordination, makes arthroscopy more ergonomics challenging task, and increases the risk of unintended tissue damage. The developed experimental stereo arthroscope has the potential to overcome these drawbacks. There are two miniaturized camera sensor boards (muC103A) placed on the camera tip. The diameter of the camera tip is about 6 mm diameter, slightly larger than standard Stryker arthroscope which is 4.0 mm. The surgical sites are illuminated by white LED (T0402W) placed on the tip of the camera. Details of the camera model can be found in [4,12].

This study includes four cadaveric knee arthroscopic video frames. The arthroscopic sequences were performed at the Medical and Engineering Research Facility (MERF) which includes four cadaver knee samples of both female and male. The incisions for tools and camera were made at the bottom left and right soft spot below the patella tendon. Water irrigation system was used to create additional space.

Apart from the challenges of the arthroscopic data, the data set obtained from our camera prototypes are limited by some factors. The dataset captured by our camera model is strongly affected by the effect of illuminations, several disturbances such as debris, noises, saturation, blur and in some occurrences out of focus. Additionally, muC103A camera board provides low resolution video frames. Moreover, to meet the surgical flow during the camera incision and in some situation during the arthroscopic course, additional light source (Xenon) was used which produced different color temperature. Due to damage, aging and degenerations, every knee structure is considered unique to patients. However, error introduced in data acquisition procedure and poor imaging conditions during camera development phase the cadaveric dataset exhibits strong discontinuity related to their appearances among the experiments. Only the last cadaver exhibits relatively high quality of video frames. Moreover, there are slightly discrepancy of view angle also observed among the cadaveric experiments, mostly due to the camera metallic shift rod was not comfortable to move. Therefore, the structural diversity is common among the dataset.

## B. Model

### A. Multi-Scale Fully Dense Connection (MS-FDC):

The elegant architecture proposed in [17] incorporates the residual learning strategy to solve degradation and vanishing gradient problems through the use of identity map [18] established via residual connections. Let's assume that, $X_l$ is the input feature map of previous layer and $X_{l+1}$ is the output feature map of current layer then, the residual convolution block is defined by the Eq. (1);

$$X_{l+1} = \sigma(X_l + F\ (X_l, W_l)) \qquad (1)$$

where F(.) known as a residual function as shown in Fig.2. $\sigma$ denotes the activation functions. $W_l$ denotes the learnable parameters of convolutional blocks.

This propagation strategy carries out information from the previous layer to the next layer. This contributes to relieving the vanishing gradient problem during training described above, therefore, in this study the proposed architecture is built on top of Deep ResUnet [17] - the architecture which combines residual learning strategies on top of UNet. Another advantage of residual learning in the context of surgical scene segmentation is that it reduces loss of low-level details, therefore, provides finer details to the high-level semantic feature maps. The other fundamental reason to use residual learning strategy is that surgical scenes exhibit very limited features and it is common to all MIS procedures. Hence, lossless propagation strategy is more desirable.

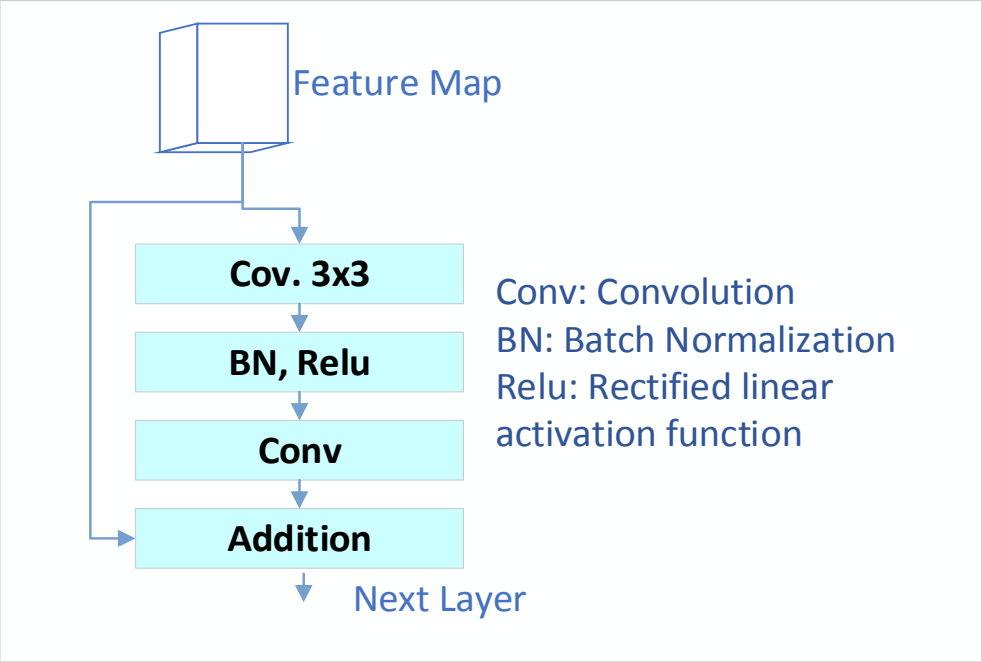

**Fig.2.** Residual connection. These are the building blocks of the proposed UNet architecture

Long range dependency between pixel to semantic map is addressed through the use of multi-scale stacked convolutions in UNet architecture. However, the stacked convolutional layer under the local neighborhood seems inefficient [19]. There are several advancements achieved to address this problem, for instance, non-local networks, attention mechanisms and multi-scale dense UNet [19-21]. Moreover, several approaches such as pyramidal image [22], atrous or dilated convolution [23], inception network [24], and spatial pyramid pooling [25] are widely used to extract multi-scale features with variable receptive fields. In their work [21], the dense connections among the layers are proposed which strengthen feature propagation and enable better gradient flow. Instead of multi-dense, we followed a fully dense connection strategy which yielded better results as explained in the result section. Therefore, we called it a multi-scale fully dense connection. We directly fuse all previous layer's outputs (fully dense) which make rich feature propagation among the encoder and decoder layers, and between the encoder to decoder. These enabled lossless multi-scale local neighborhood feature maps propagation, thus it is also effective to tackle long-short term dependency problems.

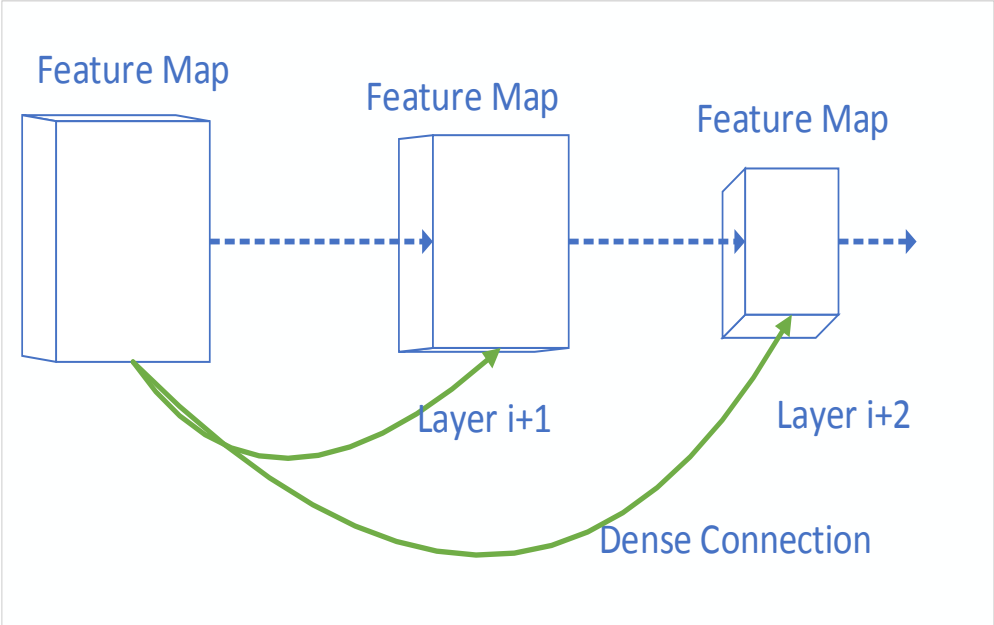

**Fig.3.** Multi-scale dense connections. These establish multiscale feature propagation among all layers.

## B. Spatial-Channel Attention Mechanism (SC-Attn):

Learnable attention mechanisms assign a score to each feature with the aim to model important ones where they suppress others. Additionally, it helps the network to model short- and long-range dependency. There are several attention mechanisms used so far with FCN, however, most of the efforts are dedicated to attaining spatial attention maps. Some others work focused channel-wise attention maps and affords recalibrated feature maps, for instance, SEnet [26] and GCnet [27]. In their work [28], J.C et al. embedded both spatial and channel wise attention mechanisms. In Arthroscopic scenes, the tissue signature also exhibits channel wise dependency, hence, instead of spatial attention gates, both spatial and channel wise attention mechanisms are more reasonable to our concern problem discussed in this study [11]. Therefore, we adapt their proposed attention mechanism into our proposed model to obtain both spatial and channel attention maps.

Attention blocks consist of two design units; namely, i.) Spatial Attention Block, and ii.) Channel Attention Block. Spatial attention block first aggregates channel information using two different pooling mechanism- average polling and max pooling. The same resolution feature maps obtained from channel wise pooling operation are then fused together through concatenation operation. Feature scores are then learned through the use of sigmoid nonlinear activation function followed by standard convolution. The resultant feature scores then multiplied with the input feature maps.

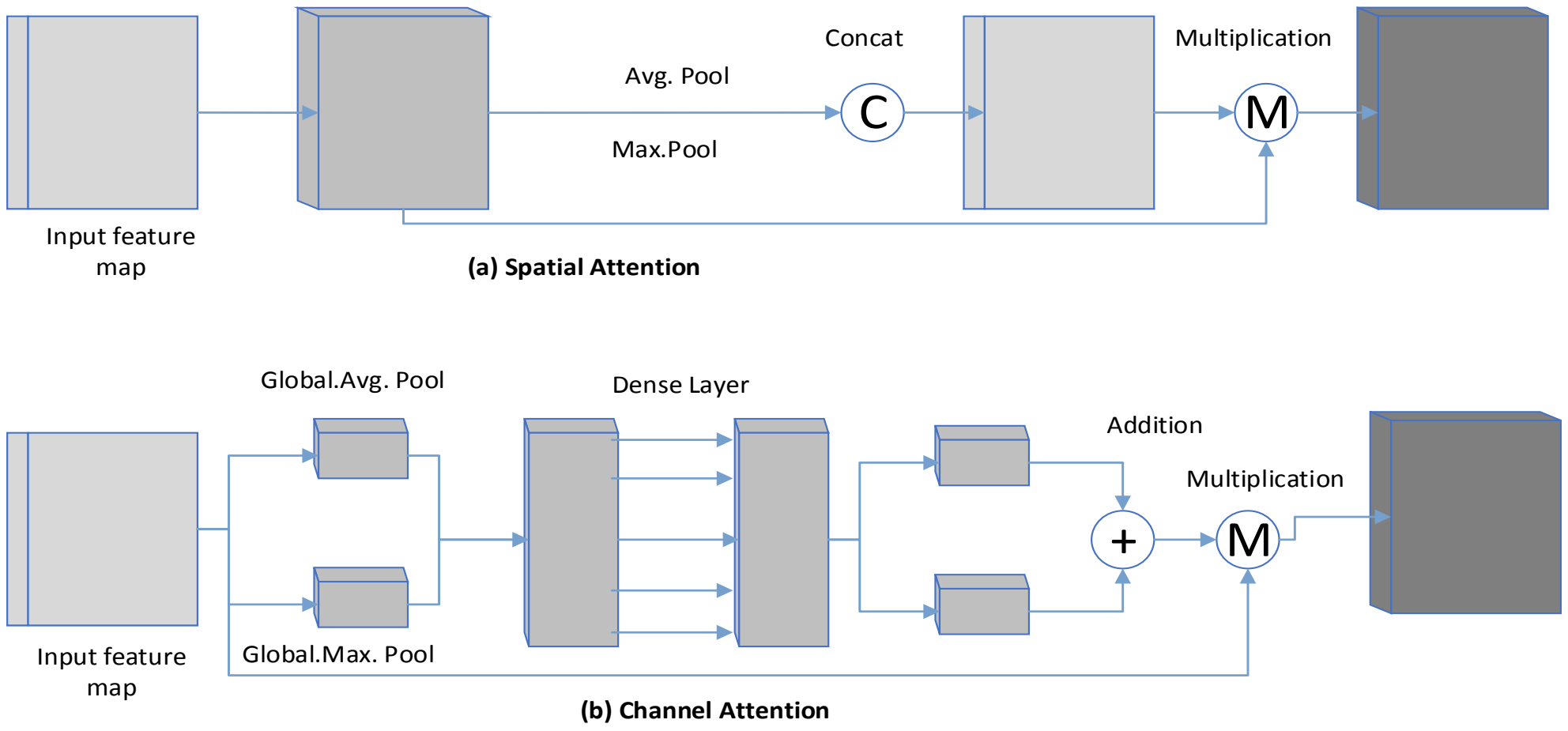

**Fig.4.** Represents Attention block. There are two attention blocks, **i.**) Spatial attention block and **ii.**) Channel attention block.

Similarly, inter channel relationships are modeled through the channel attention map. The input feature map passes through two global pooling operations - global average pooling and global maximum pooling. The resultant feature maps obtained from these pooling operations are then passed through two fully connected layers and then added together to attain a secondary channel attention map. This map then passes through a sigmoid function and the resultant score map then multiplied together with input feature map to attain channel attention map.

Where, $Conv$(.) is the convolution operation, $Cat$(.) is the concatenation operation, $Mul$(.) represents multiplication, $Add$ (.) represents addition, and $Pool$ (.) is the pooling operation. $X_{Avg}^{1xHxW}$ and $X_{Max}^{1xHxW}$ represent feature maps obtained from channel wise average pooling and maximum pooling where $X_i^{CxHxW}$ is the input feature map.

The attention map then modeled through summing up both $F_{CH_score}^{Cx1x1}$ and $F_{score}$ that pass through an additional convolution.

## C. Multi Scale Feature Extraction (MC-FE):

Multi-scale features can be extracted by Atrous Spatial Pyramid Pooling (ASPP) and Inception Block [23-24]. In ASPP, atrous convolution extracts features from various receptive fields while keeping the trainable parameter fixed. ASPP is used in this architecture at the end of the bridge layer. It extracts multi-scale features from high level features map of the encoder side and creates a four-scale feature pyramid. Moreover, maximum pooling of size equal to the size of the feature map extract high level global contextual information. The signals are then concatenated together which propagates features along with contextual information to the decoder block to reduce semantic gaps.

On the other side, the inception block is used at the end of the dense decoder blocks which fuses all the multi-scale futures. In the inception block, the multi-scale features are extracted through the use of various kernel sizes as depicted in Fig. 6**.** The basic difference between ASPP and inception block here, it does not include global pooling but a maximum pooling with size 3x3 which brings salient features to the resultant concatenated feature map.

In FCN, the global context retrieves through repetitive convolutional operation in a local window and henceforward, the spatial and channel wise information are fused together to represent complete feature maps [26]. In their work, the squeeze and excitation blocks are proposed with the aim to capture global contextual information through global pooling and establish channel-wise recalibrated feature maps [26]. With the aim of propagating recalibrated global contextual information from high features to the semantic layers, a global context block is used at the end of the bridge layer.

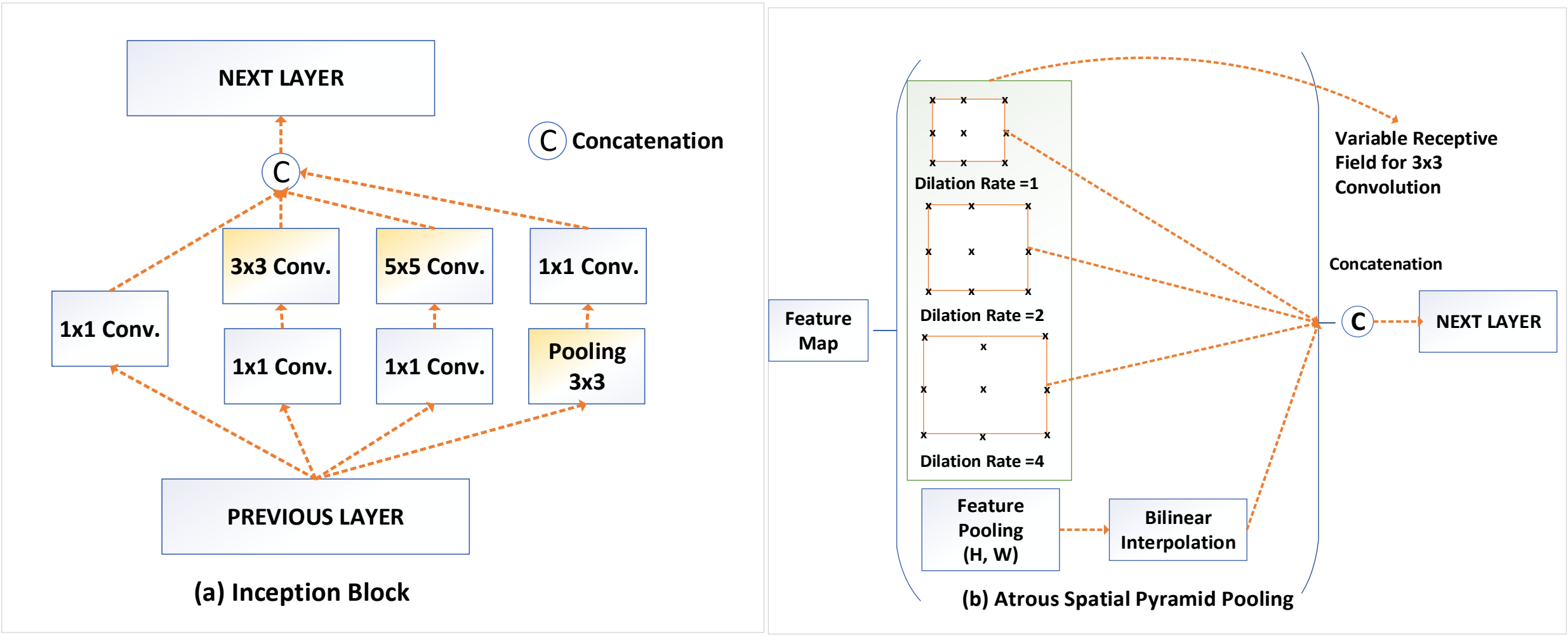

**Fig.5.** represents Inception block and SPP with atrous convolution to extract multi-scale features.

## D. Shape Extractor Block (SEB):

Surgical scenes exhibit few features and highly texture less areas. In some circumstances though random texture exists due to tissue degenerations but often they are not robust and enough to represent unique tissue structures, nevertheless, these diverse natures of the tissue appearances subsequently exhibit high intra-tissue discrepancy. However, it has been observed that the shape features can improve this dilemma and it is also confirmed that to some extent biomedical image analysis, the shape features exhibit robust visual cues, for instance, the cardiac MRI scan.
In this study, we use a minimalist implementation of the FCN architecture - UNet to extract the shape of the key structures. This branch of subnetwork is added to the segmentation network as shape extractor. The aim of the shape extractor branch to model the shape of the tissue structure of ACL, Meniscus, Femur, Tibia, Patella and Tool.
Though ACL, Meniscus, Bones (Femur and Tibia) and tools do not exhibit enough discriminative local features among themselves, they are substantially different from other unstructured tissue types e.g. fat or floating tissue. Moreover, local boundary information is also an indispensable information to model their shapes with the presence of strong edges.
Following the standard UNet architecture, the shape extractor consists of three branches, namely i.) contracting path, ii.) bridges, iii) expansion path. Global context used at the end of the bridge to extract global contextual information. The institution is to model shape from local features; therefore, only spatial attention blocks are used with the skip connections which can propagate important spatial features. To establish rich information propagation with multi-scale semantic maps, all decoder layer outputs are concatenated before the final layer. At the final layer sigmoid activation function is used that produces a binary shape-aware segmented map.

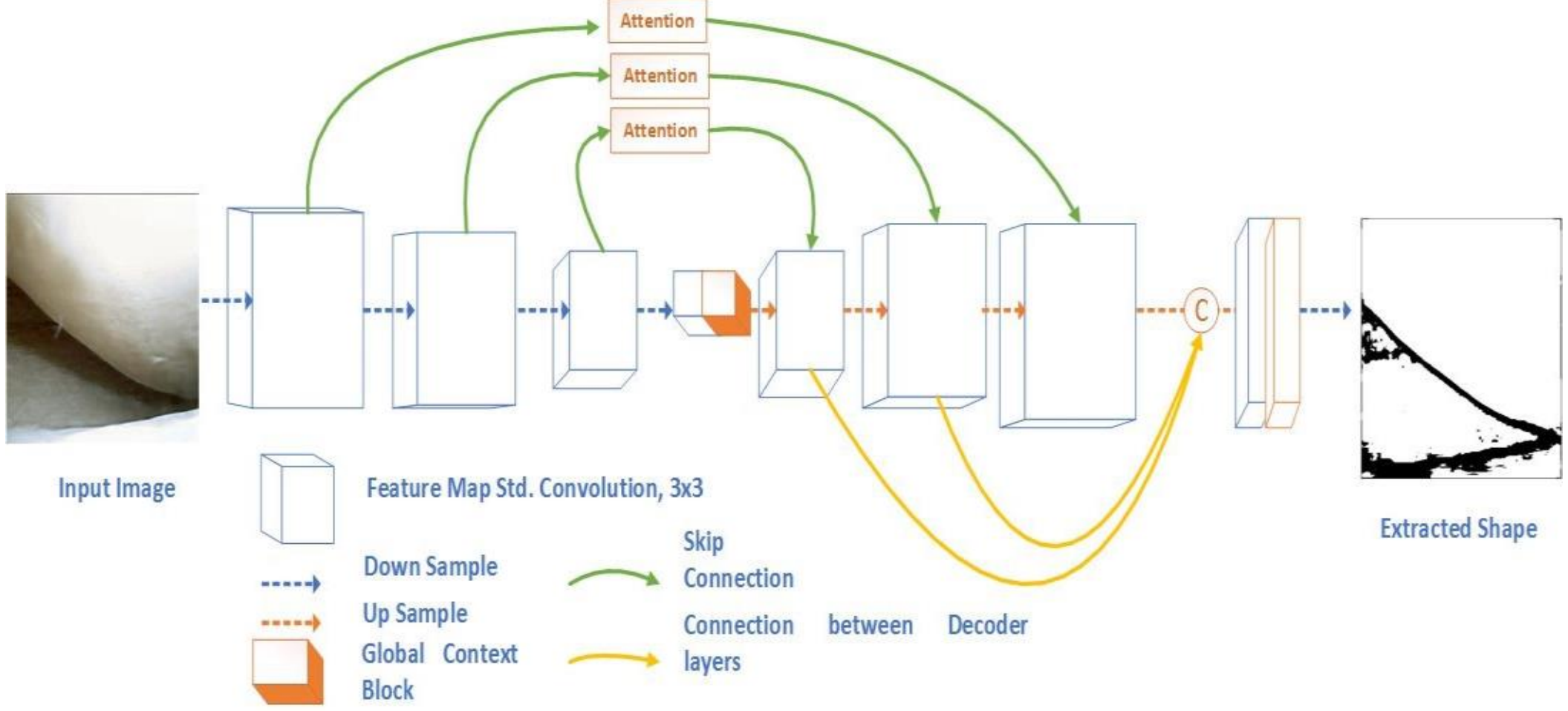

**Fig.6:** Architecture of the shape extractor block. Global context block and short connections between the decoder layers are used with this minimal implementation of UNet that ensures rich information propagation. The network provides shape features for the key structures.

Each layer consists of two consecutives standard 3x3 convolution and rectified linear activation function (ReLU). At the end of the bridge layer a global context block is used. The spatial attention blocks used to skip connections between the encoder and decoder. Moreover, the decoder path consists of two dense connections that enrich feature propagation and brings low level shape information to high level shape aware semantic features. All the semantic features are then concatenated which is then followed by standard 3x3 convolution. At the last layer,the dimension of the segmented map is adjusted by 1x1 convolution.

## iii. Network Implementation

**Fig.7.** represents the overall architecture of our proposed network. The TensorFlow library is used to implement the entire network. The shape extractor sub-network can be trained together with the model or as a standalone network. We trained the shape extractor sub-network separately for each four-fold-validation and then deployed the pretrained network. The aim of the shape extractor is to capture the shape information of the key tissue structures of interest, in this study it was the femur, tibia, meniscus, ACL, patella, and any surgical tools present. The shape extractor also discriminates unstructured features such as fat and dense collagen as background. The accuracy of the shape extractor was relatively high due to the tissue structures of interest showing few discriminative texture and features, but they are significantly different from unstructured tissue types. The average dice similarity coefficient for 4-fold-validation was 91.78%. The dice similarity coefficient measures similarity between the ground truth and predicted label for each class type which is defined in Eq (10).

The shape feature map is rolled back to the input layer of the segmentation network to facilitate a shape-aware spatial feature learning mechanism. The convolution block of each layer of the segmentation network is built upon residual connection strategy. The down sample is performed through stride convolution rather than pooling operation to enable minimum information loss compared to pooling operations [30].

Each layer’s input contains a concatenated feature map of all previous layer outputs, then a 1x1 convolution is performed to adjust filter size, followed by a standard 3x3 convolution. The resultant feature map then passes through the residual block. The output of the residual block then has two connections, one that connects to the subsequent layers and another which is densely connected to the decoder layers.

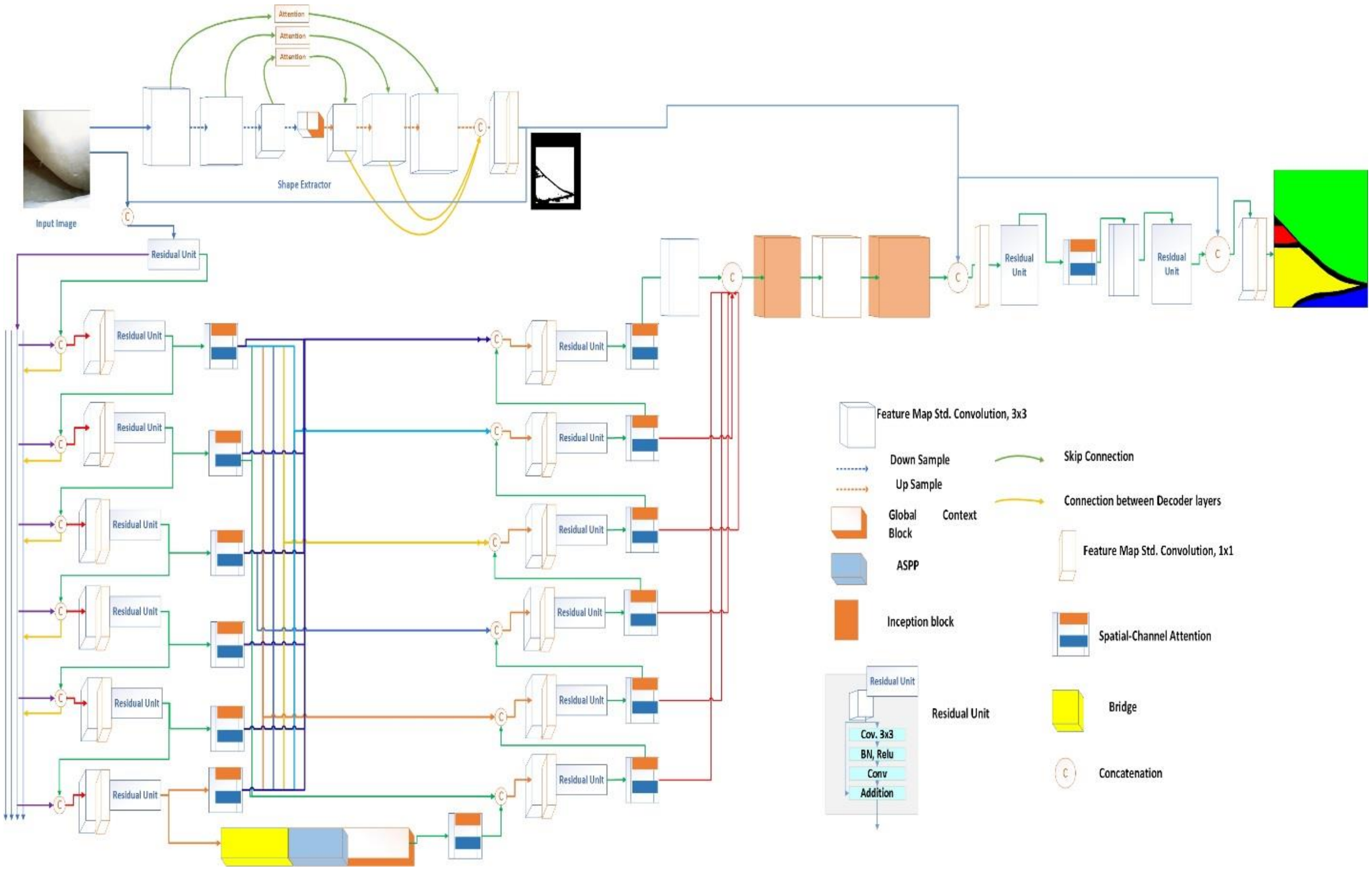

**Fig.7.** Implementation of segmentation network. Segmentation network contains two basic branches, i.) shape extractor block and, ii.) main segmentation network. Upper branch of the complete network is the shape extractor block which takes the surgical frame as an input and extracts the shapes of the key tissue structures. The shape features are (binary mask contains tissue shapes) rolled backed and concatenated with the input surgical scene frame. Therefore, shape mask and scene frame are the inputs of segmentation network block and it enables shape aware feature extraction mechanism. Each network layer of segmentation block contains dense connection among the others. In bridge layer between the encoder and decoder, global context and ASPP blocks are used to capture multi-scale global information from high level feature maps. During the final stages, inception blocks, fusion of multi-scale semantic maps, and shape feature are used obtain semantic segmented map.

The bridge layer contains two standard convolutions, ASPP block is then used, thereby producing a multi-scale high level feature map. A global context block is then used to extract a high level rich global contextual channel-wise recalibrated feature map. The resultant feature map then passes to the decoder block.

Each layer of the decoder block has a dense skip connection to receive multi-scale channel and spatially aware feature maps from the encoder side. Dense skip connections enrich information propagation and maintain long-term dependency. Furthermore, after the final decoder layer, all previous layer outputs from the decoder are concatenated to propagate multi-scale semantic features. After this, two inception blocks which extract semantic features with a variable receptive field are used. Inception blocks provide a large set of features map through concatenation operations. Therefore, the dimension conversion between each inception block is adjusted through 1x1 convolution to learn salient feature mapping.

The shape feature is then concatenated on top of the semantic feature maps obtained from the inception blocks. Two residual convolutions followed by an attention mechanism then produce the shape aware semantic feature map before performing final labeling. The final layer consists of one 1x1 convolution with filter size 3. The Sigmoid activation function is used to achieve a probabilistic value for each channel.

Filter sizes used in our implementation are 8, 16, 32, 64, 128, and 512. The network is optimized using Adam optimizer. The initial learning rate is set to 1e-4 and uses a keras learning scheduler with decay rate 0.9. The network is trained with combined loss function of categorical cross entropy and dice coefficient (DCE) as discussed in [1, 31]. The total loss is defined as;

$$DCE\ (T,P) = 0.5 * CCE(T,P) + (1 - Dice(T,P)) \qquad (2)$$

Where Dice loss can be calculated as follows;

$$Dice(T,P) = (2 * \sum T_i\, P_i + Smooth)/(\sum T_i + \sum P_i + Smooth) \qquad (3)$$

Where, $T_i$ is the ground truth label and $P_i$ is the predicted label. Smooth is the smoothing factor typically set to 1. The categorical cross entropy is defined as;

$$CCE(T,P) = (\frac{1}{N}) \sum_{i}^{N} \sum_{j}^{C} T_{ij} \log P_{ij} \quad (11)$$

## IV. Experimental Results

We conducted the four-fold cross-validation experiment for the dataset obtained from our stereo camera. Based on this, we used arthroscopic data from three cadavers to train our network and an arthroscopic dataset from one cadaver to validate our model. To remove artifacts, the video frames were pre-processed following the method proposed by Shahnewaz Ali [14]. We used a total of 3800 video frames. The data augmentation mechanisms applied to all frames were shift and rotation, and vertical and horizontal flip using Albumentations [32]. To tackle class imbalance problems, we reused lower frequency tissue-structure with additional data augmentation mechanisms; namely, scale-shift, center-crop, and zoom-in so that the network could learn. However, in many frames there were small fractions of ACL, meniscus, and tools present compared to other tissue types in the same frame. In most situations, it was femur. We selected the frames and, in some situations, augmented (cropped) them in such a way that femur exposure remained minimal. Table-I shows the cross-validation results for the stereo camera.

**TABLE I**. CROSS VALIDATION RESULT OF STEREO DATASET

| Validation Set | ACL | Femur | Tibia | Meniscus | Tool and Patella |
|---|---|---|---|---|---|
| Dataset 1 | 0.701 | 0.888 | 0.682 | 0.430 | 0.989 |
| Dataset 2 | 0.20 | 0.790 | 0.676 | 0.290 | 0.931 |
| Dataset 3 | 0.337 | 0.942 | 0.782 | 0.715 | 0.623 |
| Dataset 4 | 0.7894 | 0.9810 | 0.662 | 0.686 | 0.352 |
| **Average** | **0.506** | **0.900** | **0.7005** | **0.530** | **0.7235** |

It is worthwhile to mention that we received relatively high quality of images from the first and last cadaveric experiment. In the first cadaver experiment, tissue type ACL, bone cartilage and meniscus were found in a healthy state, which had similarity with the last cadaver. Hence, cadaver dataset one and four received high validation accuracy.

Cadaver experiment two and three (datasets 2 and 3) had strong cartilage degeneration and relatively poor visual cues for tissue type ACL. Moreover, these two cadaveric datasets were highly affected by the illumination conditions. Many frames were extremely oversaturated or underexposed. Therefore, ACL tissue structure received relatively lower accuracy. However, the proposed network efficiently recognizes degenerated cartilage.

Due to the camera pose, view and incision angles, in the four-cadaver data there is major discrepancy in the appearance of frames. This was due to different view angles in some frames the representation of the tissue arrangement of meniscus, tibia, ACL and femur depict high discrepancy among the other dataset. Mostly tissue structure meniscus and in some frames, ACL suffer from this limitation. It is meaningful to mention that, due to the stereo camera supporting wider FoV, the data inconsistency is more observable. In order to tackle inter-patient variability and discrepancy caused by camera view angle, we used 50 frames from 1125 frames to train our network which represent completely unseen structural appearance among other datasets.

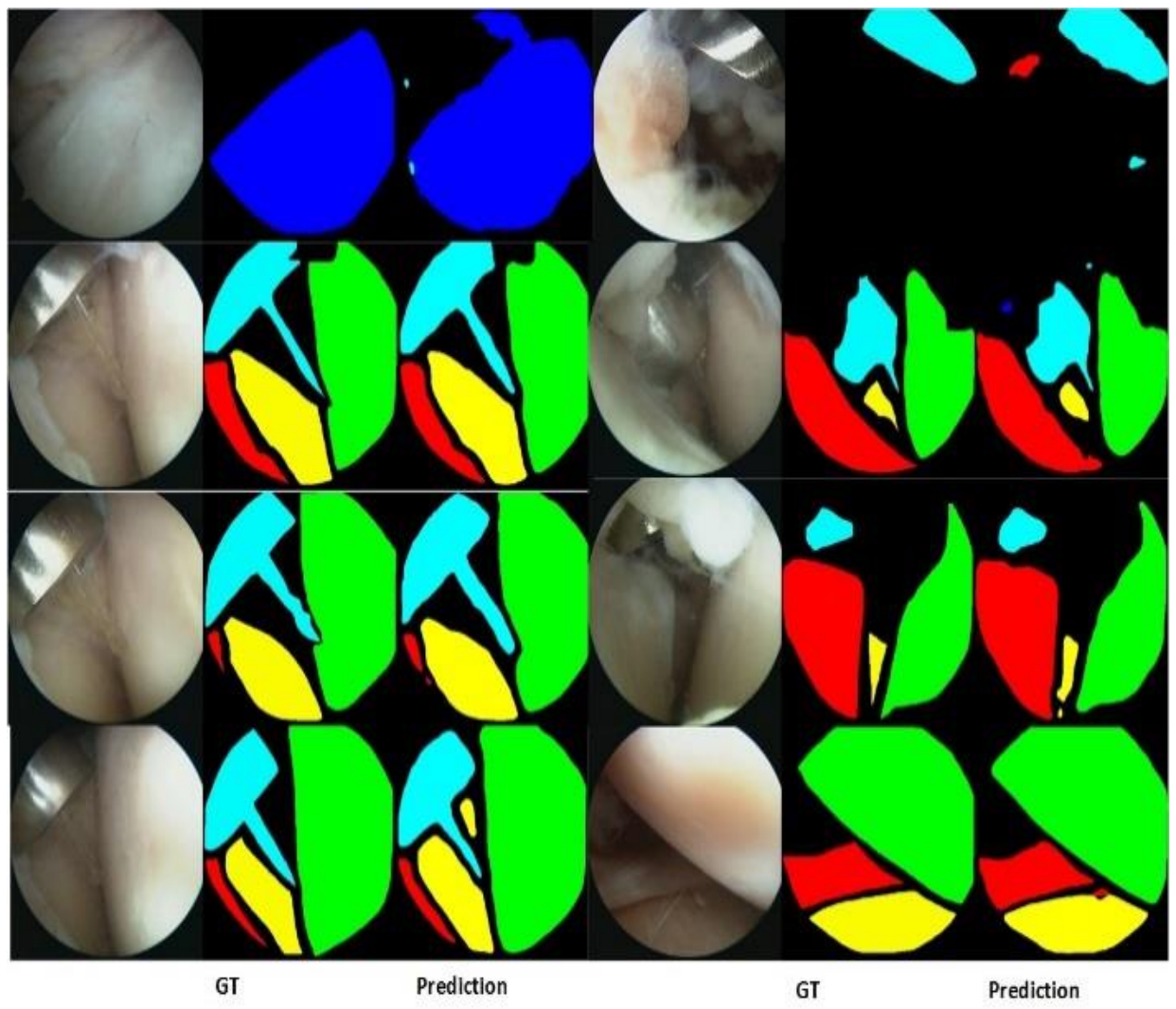

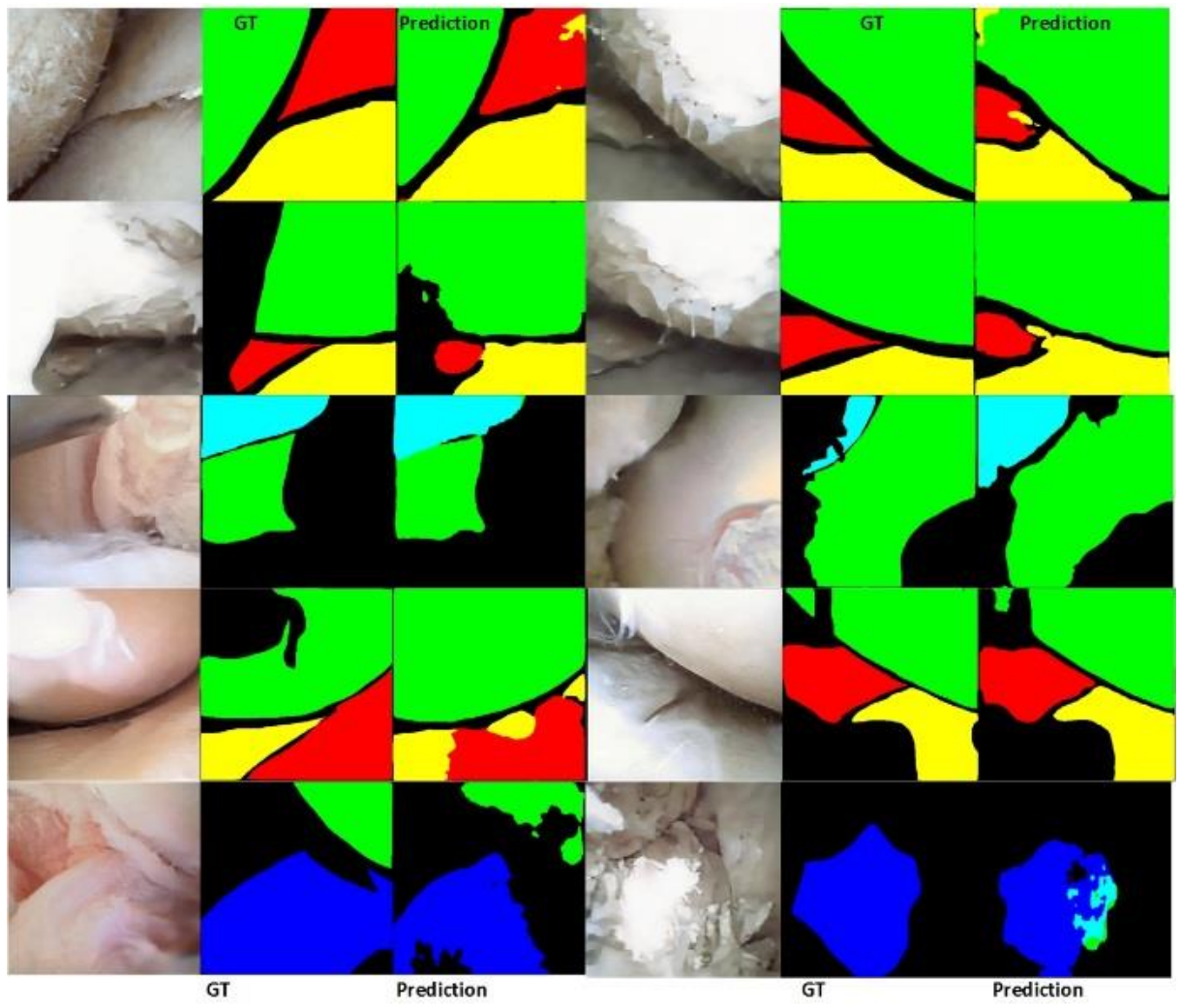

(a) Stryker camera (b) Our Camera

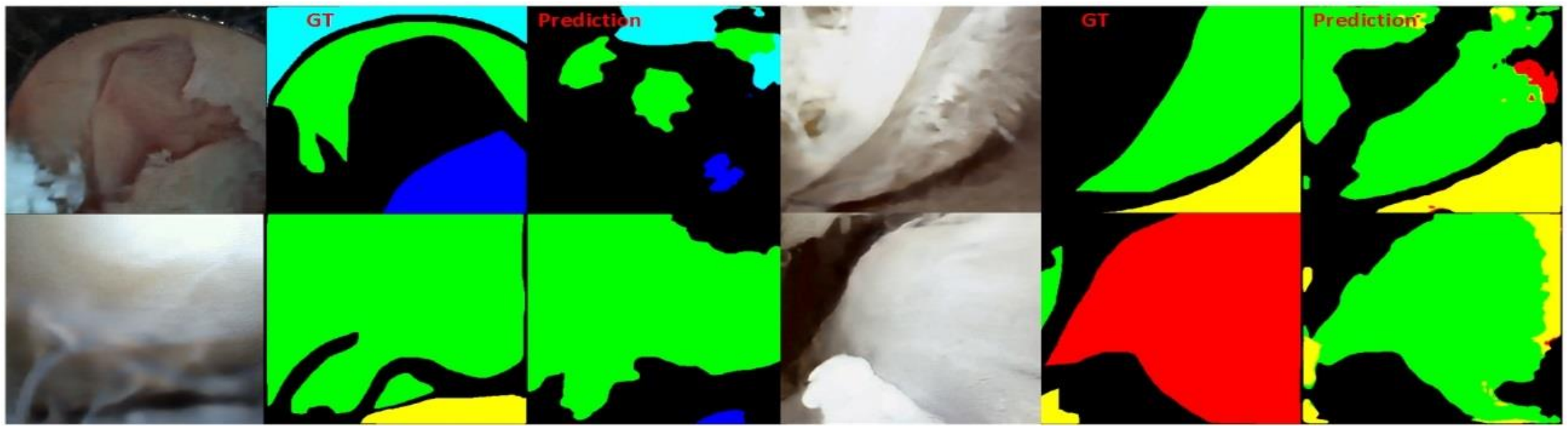

(c) Poor segmentation

**Fig.8.** Segmented maps of arthroscopic surgical scenes obtained from our proposed method. (a) represents segmentation result for Stryker endoscope while (b) represents video frames captured from our developed stereo endoscope. (c) represents relatively poor segmentation result achieved due to poor imaging condition and lack of visual information such as shape.

We also validated our model with a publicly available dataset to compare model efficacy against three state-of-the-art models namely; ResUNet++, ResUNet, and UNet. In their article [33], the result is presented for the task of Polyp segmentation from colonoscopy dataset. Kvasir-SEG [34] dataset is used which contains 1000 endoscopic image and ground-truth annotations by domain experts of Oslo University Hospital (Norway). In their experiment, they used 80% data for training, 10% of data for validation and 10% of data for testing. Dice coefficient loss was used as the loss function along with Adam optimizer with a 1e-4 learning rate. To compare our model in the most direct manner we used the same training settings. As the table shows, our model achieved relatively high accuracy with 5.09% accuracy improvement.

TABLE III **TEST RESULT – POLYP SEGMENTATION**

| Model Name | Accuracy (Dice Coefficient) |
|---|---|
| ResUNet++ | 81.33 % |
| ResUNet-mod | 79.09 % |
| UNet | 71.47 % |
| **Ours** | **86.42 %** |

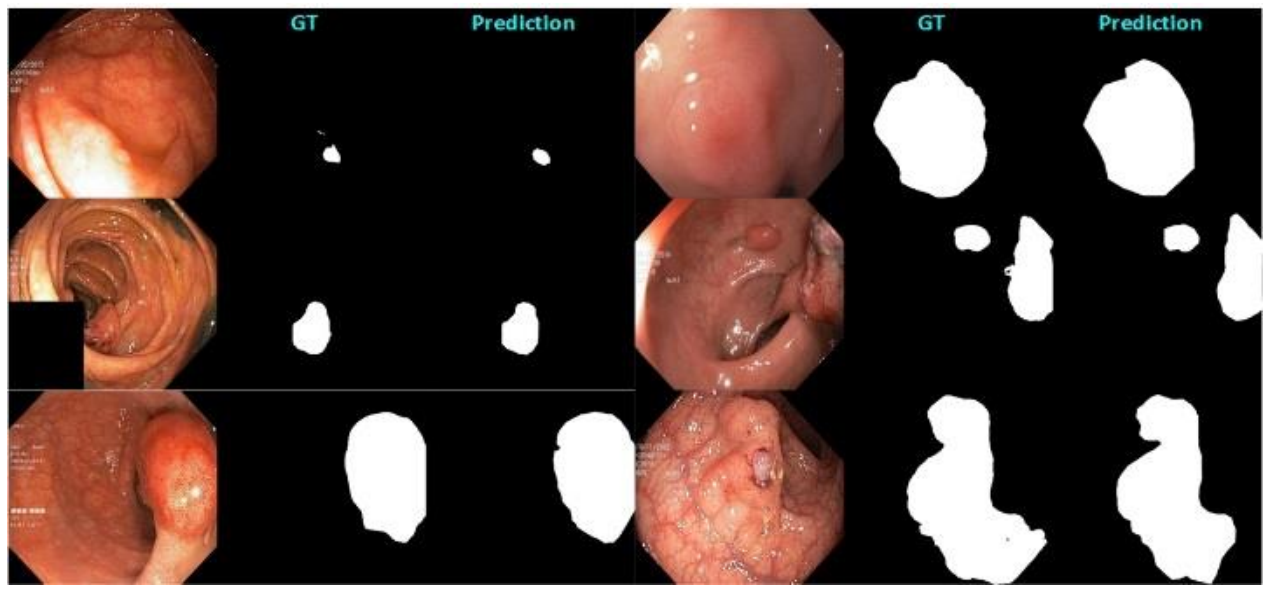

**Fig-9**. Polyp Segmentation result.

Table IV represents compressive qualitative results when different design blocks are added to our model. Model 1 represents ResUnet with shape extractor block, model 2 represents ResUnet with shape, model 3 represents model 2 with multiscale feature extractor namely ASPP block, model 4 represents model 3 with dense multi-scale feature propagation and inception block, and finally model 5 represents model 4 with shape aware spatial feature learning mechanism (shape feature and image as an input).

TABLE IV **MODEL EVALUATION**

| Model Name | Unit ResUnet | Shape Extractor | Multiscale feature (ASPP) | Multiscale feature (ASPP+ Inception) | Dense multi-scale feature | Shape aware spatial feature learning | Accuracy (Dice Coefficient) |
|---|---|---|---|---|---|---|---|
| Model 1 | ✓ | ✗ | ✗ | ✗ | ✗ | ✗ | 69% |
| Model 2 | ✓ | ✓ | ✗ | ✗ | ✗ | ✗ | 70.03% |
| Model 3 | ✓ | ✓ | ✓ | ✗ | ✗ | ✗ | 71.82% |
| Model 4 | ✓ | ✓ | ✓ | ✓ | ✓ |  | 75.65% |
| **Model 5** | ✓ | ✓ | ✓ | ✓ | ✓ | ✓ | **78.14%** |
| ResUnet++ | ✓ |  | ✓ | $2^{nd}$ ASPP | SE blocks |  | 72.55% |

## V. Conclusion

In this paper, we propose a deep learning model for arthroscopy to achieve automatic segmented maps of key anatomical structures. Our model successfully creates segmented maps for the five key structures namely: ACL, Femur, Tibia, Meniscus, Patella and Tools present in knee arthroscopic video frames with average dice score 0.506, 0.9, 0.7, 0.53, and 0.72. For autonomous robot-assisted knee arthroscopy it has great impact to navigate and localize camera and tools safety. Moreover, in a confined space like knee bone-joint, if the key structures are recognized it will assist to cancel out the risk of unintentional tissue damage. Additionally, the model can help to train new clinicians.

Our model is validated and tested with three different datasets. Relatively low accuracy is achieved when the model is cross validated with the stereo dataset. Main reasons are the data discrepancy, poor imaging and lighting conditions, as well as errors introduced during video acquisition. Moreover, it is more desirable to have enough representative video frames to capture inter-patients knee structural variability. However, despite these drawbacks our proposed model achieved relatively higher accuracy. Significant accuracy improvement is received when the model is tested with the conventional imaging device. Moreover, we tested our model with publicly available polyp segmentation dataset and received highest accuracy among others.

### ACKNOWLEDGMENT

This work is supported by Australian Indian Strategic Research Fund (AISRF), The Medical Engineering Research Facility and QUT Centre for Robotics.